\documentclass[preprint2]{proto}

\pdfoutput=1

\usepackage{graphicx}
\usepackage{times}
\def\asec{$^{\prime\prime}~$}

\newcommand{\refs}{\par\noindent\hangindent=1pc\hangafter=1}
\begin{document}

\title{\textbf{\LARGE Multi-Wavelength Imaging of Young Stellar Object 
Disks:\\ Toward an Understanding of Disk Structure and Dust Evolution}}

\author {\textbf{\large Alan M. Watson}}
\affil{Universidad Nacional Aut\'onoma de M\'exico}

\author {\textbf{\large Karl R. Stapelfeldt}}
\affil{ Jet Propulsion Laboratory, California Institute of Technology}

\author {\textbf{\large Kenneth Wood}}
\affil{University of St. Andrews}

\author {\textbf{\large Fran\c cois M\'enard}}
\affil{Laboratoire d'Astrophysique de Grenoble} 

\begin{abstract}
\baselineskip = 11pt
\leftskip = 0.65in
\rightskip = 0.65in
\parindent=1pc

 
{\small We review recent progress in high-resolution imaging of
  scattered light from disks around young stellar objects. Many new disks have
  been discovered or imaged in scattered light, and improved
  instrumentation and observing techniques have led to better disk
  images at optical, near-infrared, and thermal-infrared wavelengths.
  Multi-wavelength datasets are particularly valuable, as dust particle
  properties have wavelength dependencies. Modeling the changes in
  scattered-light images with wavelength gives direct information on the
  dust properties. This has now been done for several different disks.
  The results indicate that modest grain growth has taken place in some
  of these systems. Scattered-light images also provide useful
  constraints on the disk structure, especially when combined with
  long-wavelength SEDs. There are tentative suggestions in some disks
  that the dust may have begun to settle. The next few years should see
  this work extended to many more disks; this will clarify our
  understanding of the evolution of protoplanetary dust and
  disks.\\~\\~\\~}
 
\end{abstract}

\section{INTRODUCTION}

\begin{figure*}[t]
\epsscale{2.0}
 \includegraphics[width=\linewidth]{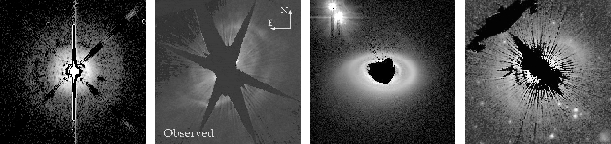} 
\caption{\small Scattered light images of four face-on or intermediate inclination YSO disks newly resolved
  since 1998. Left-to-right: TW Hya face-on disk ({\it Krist et
  al.\@}, 2000), HD 100546 ({\it Grady et al.\@}, 2001), HD 141569A ({\it Clampin 
  et al.\@}, 2003), and HD 163296 ({\it Grady et al.\@}, 2000).}
\end{figure*}

This chapter reviews the progress since the {\it Protostars and Planets
IV} meeting ({\it McCaughrean et al.\@}, 2000) in observations and modeling
of scattered-light images of disks around young stellar objects (YSOs).
For many years these disks were studied in the infrared and at
millimeter wavelengths, but without the spatial resolution necessary to
reveal their detailed structure. With the advent of the {\it Hubble
Space Telescope} (HST) and ground-based adaptive optics (AO) systems,
these disks were seen in scattered optical and near-infrared starlight.
The first images of disks in scattered light showed them as never
before. These images currently provide the highest spatial resolution
images of disks and provide unique information on disk structure and
dust properties.

Most disks around young low- and intermediate-mass stars fall into
one of two categories on the basis of their gross observational
properties. YSO disks are optically thick at visible and near-infrared
wavelengths, are rich in molecular gas, and are found around Class I and
Class II systems. Debris disks are optically thin at optical and
near-infrared wavelengths, have only trace quantities of gas, and are
found around Class III and older systems.

Our current understanding of the evolution of dust in disks around young
low- and intermediate-mass stars is that these observational categories
correspond to very distinct phases. In YSO disks our understanding is
that gas and dust from molecular cores is processed through the disk and
provides the raw material both for accretion and outflows. The dust is
processed first in the dense molecular core and then in the disk itself.
The dust grows and suffers chemical processing. Dust growth is hypothesized
to result in the production of planetesimals, which form rocky planets and 
the cores of gas giant planets. In debris disks, our understanding is that 
dust is present largely as a result of collisions between planetesimals. 
Thus, YSO disks are thought to be characterized by dust growth whereas 
debris disks are thought to characterized by planetesimal destruction.

In this chapter we focus on observations and modeling of scattered-light
images of YSO disks around low- and intermediate-mass stars. That is, we
focus on the phase in which dust is expected to grow. Of course,
scattered-light observations and models are not the only means to
explore these disks. The chapter by {\it Dutrey et al.\@} discusses
millimeter and sub-millimeter observations of the gas and dust
components and the chapters by {\it Bouvier et al.\@}, {\it Najita et
al.\@}, {\it Millan-Gabet et al.\@} cover observations of the inner disk
region, notably interferometry, and their interpretations. The chapter
by {\it Monin et al.\@} discusses aspects of disks that are peculiar to
binaries. Further from the subject of our chapter, the chapter by {\it
Cesaroni et al.\@} discusses disks around young high-mass stars and the
chapter by {\it Meyer et al.\@} reviews debris disks around solar-type
stars.

In the following sections we summarize the data available on
scattered-light disks around YSOs, imaging techniques, and what we can
learn about disk structure and dust properties from modeling
multi-wavelength images and SEDs. We also speculate about future
advances.

                                                                 
\section{OBSERVATIONAL PROGRESS}

\begin{figure*}[t]
\epsscale{2.0}
 \includegraphics[width=\linewidth]{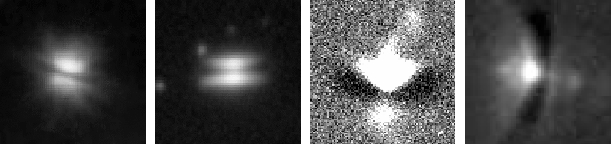} 
\caption{\small Scattered light images of four edge-on or silhouette YSO
  disks newly resolved since 1998. Left-to-right: Edge-on disks of HV
  Tau C ({\it Stapelfeldt et al.\@}, 2003), 2MASSI1628137-243139 ({\it
  Grosso et al.\@}, 2003), IRAS 04158+2805 ({\it M\'enard et al.\@}, in
  preparation), and Orion 216-0939 ({\it Smith et al.\@}, 2005).}
\end{figure*}

At the time of the {\it Protostars and Planets IV} meeting, held in
1998, only a dozen YSO disks had been resolved in scattered light and
another half dozen in silhouette against the Orion nebula. Since then,
the number of resolved disks in these two categories has doubled and
quadrupled, respectively. The new discoveries stem largely from targeted
high-contrast imaging of greater numbers of YSOs and wide-field imaging
surveys of star-forming clouds. In the optical and near-infrared, the
broader application of established observing techniques, rather than any
instrumental improvements, has driven the recent expansion in the number
of disks discovered. At longer mid-infrared wavelengths, larger
telescopes equipped with more potent mid-infrared imagers became
available. These results are described in this section and images of
some of the new disks are shown in Figs.~1 and 2. A complete list of
published YSO disks resolved in scattered light appears in Table~1. Most
of these are around T~Tau stars, although some are around Herbig AeBe
stars. An on-line catalog of circumstellar disks resolved in scattered
light, thermal emission, or molecular lines is now available; the URL is
http://www.circumstellardisks.org.

\subsection{Optical and Near-Infrared Imaging}

\noindent
{\em 2.1.1 Coronagraphic and Direct Imaging.} The most general case, and
also the most observationally challenging, is the detection of disk
scattered light in the presence of direct starlight. Subtraction of a
fiducial reference star is almost always necessary to reveal the disk,
and a coronagraph is usually employed to suppress stellar diffraction.
Various HST instruments have produced excellent images of disks around
bright stars, while groundbased AO imaging of these disks is often
hampered by the instability of the point spread function.

The brightness of Herbig AeBe stars makes them excellent disk imaging targets,
and three have been found to show interesting circumstellar nebulosity.
HD 100546 has a disk viewed from high latitudes which contains what
appear to be two spiral arms ({\it Pantin, Waelkens, and Lagage}, 2000;
{\it Grady et al.\@}, 2001; {\it Augereau et al.\@}, 2001). HD 163296 shows
an inclined disk with a hint of a cleared central zone and bifurcation
into upper and lower reflection nebulosities ({\it Grady et al.\@}, 2000).
The nebulosity around AB Aur shows a wealth of structure ({\it Grady et
al.\@}, 1999; {\it Fukagawa et al.\@}, 2004), but it is unclear how much is
associated with the disk and how much with a more extended envelope.
{\it Fukagawa et al.\@} (2003) report a possible disk around HD 150193A.

The transitional disk HD 141569A has been the subject of a great deal of
observational and theoretical work.  The first images in the
near-infrared by {\it Weinberger et al.\@} (1999) and {\it Augereau et al.\@}
(1999) appeared to show a large radial clearing in the disk at a
radius of 250 AU.  Optical imaging by {\it Mouillet et al.\@} (2001) and
{\it Clampin et al.\@} (2003) revealed an asymmetric spiral-like feature
and showed that the cleared region was not completely empty.  HD 141569A
is in a multiple system. Dynamical studies suggest that stellar fly-bys
or a recent periastron passage by the companions could be the origin of
the observed spiral feature ({\it Augereau and Papaloizou}, 2004;
{\it Ardila et al.\@}, 2005; {\it Quillen et al.\@}, 2005).

Perhaps the most significant new disk imaged around a directly visible
star since {\it Protostars and Planets IV} is that of TW Hya ({\it Krist
et al.\@}, 2000; {\it Trilling et al.\@}, 2001; {\it Weinberger et al.\@},
2002). This face-on disk is also the closest T Tauri star disk, at a
distance of 56 pc, and has been well-studied at millimeter wavelengths.
The disk radial profile shows a sharp break in slope at a radius of 140
AU which is seen in multiple independent datasets. The physical origin
of this sudden fading in the outer disk is unclear.   Long-slit spectra
of the TW Hya disk with HST STIS ({\it Roberge et al.\@}, 2005) find that 
the disk has a roughly neutral color from 5000 {\AA} to 8000 {\AA} 
(see their Fig.\ 5), consistent with previous broad-band imaging in the 
same spectral region ({\it Krist et al.\@}, 2000) and near-infrared 
({\it Weinberger et al.\@}, 2002).

Despite the interesting results highlighted above, most YSO disks
(those whose presence is inferred from infrared and submm excess
emission) remain undetectable in scattered light. There are probably
multiple reasons for this fact. Disks with outer radii smaller than
0.5\asec, the typical inner working angle cutoff for current
coronagraphs and PSF subtraction techniques, would not extend far enough
from their host star to be readily detected. This cannot be the whole
explanation, however; the disk of MWC 480 has a large outer radius in
millimeterwave CO emission ({\it Simon et al.\@}, 2000), but repeated
attempts to detect it in scattered light have failed ({\it Augereau et
al.\@}, 2001; {\it Grady et al.\@}, 2005). There are other examples. Many
disks may simply be intrinsically fainter than the ones that have been
imaged in scattered light to date; they may be geometrically flatter
(and thus tending toward self-shadowing), depleted of small particles, 
or have undergone chemical processing that has reduced their albedo. 
With the typical YSO disk currently undetectable against the 
direct light of its parent star, improved coronagraphic instrumentation 
will be needed to understand the diversity in disk scattering properties 
and their underlying causes.

\begin{deluxetable}{llrl}
\tabletypesize{\small}
\tablecolumns{1}
\tablewidth{0pt}
\tablecaption{YSO Disks Imaged in Scattered Light}
\tablehead{Object Name & Type         & Outer Radius & Recent Reference }
\startdata
TW Hya       & face-on      &     220 AU      &{\it Roberge et al.\@} (2005) \\
HD 141569    & transition   &     370 AU      &{\it Clampin et al.\@} (2003) \\
HD 100546    & spiral       &     360 AU    &{\it Grady et al.\@} (2005) \\       
HD 163296    &              &     450 AU    &{\it Grady et al.\@} (2005) \\
HD 150193A   &              &     190 AU    &{\it Fukagawa et al.\@} (2003) \\
AB Aur       &disk+envelope    & $>$ 300 AU    &{\it Fukagawa et al.\@} (2004) \\
CB 26        & edge-on      &     380 AU    &{\it Stecklum et al.\@} (2004) \\
CRBR 2422.8-3423 & edge-on  &    105 AU     &{\it Pontoppidan et al.\@} (2005) \\
DG Tau B   & edge-on      &     270 AU    &{\it Padgett et al.\@} (1999) \\
GG Tau     & CB ring      &     260 AU    &{\it Krist et al.\@} (2005) \\
GM Aur   &              &     500 AU    &{\it Schneider et al.\@} (2003) \\
HH 30        & edge-on      &     225 AU    &{\it Watson \& Stapelfeldt} (2004) \\
HK Tau B   & edge-on      &     105 AU    &{\it McCabe et al.\@} (2003) \\
HV Tau C   & edge-on      &     85 AU     &{\it Stapelfeldt et al.\@} (2003) \\
Haro 6-5B    & edge-on      &     280 AU    &{\it Padgett et al.\@} (1999) \\
IRAS 04302+2247 & edge-on   &     420 AU    &{\it Wolf et al.\@} (2003) \\ 
UY Aurigae   & CB ring      &    2100 AU    &{\it Potter et al.\@} (2000) \\
2MASSI J1628137-243139 & edge-on & 300 AU   &{\it Grosso et al.\@} (2003) \\
IRAS 04325+2402 & edge-on   &         30 AU &{\it Hartmann et al.\@} (1999) \\
OphE-MM3     & edge-on      &     105 AU    &{\it Brandner et al.\@} (2000a) \\       
ASR 41       & shadow       &    $<$ 3100 AU&{\it Hodapp et al.\@} (2004) \\
LkH$\alpha$ 263 C  & edge-on & 150 AU       &{\it Jayawardhana et al.\@} (2002) \\
Orion 114-426  & edge-on    &     620 AU    &{\it Shuping et al.\@} (2003) \\
Orion 216-0939 & edge-on    &     600 AU    &{\it Smith et al.\@} (2005) \\
\enddata
\end{deluxetable}



\noindent{\em 2.1.2 Edge-On Disks.} Edge-on, optically thick disks
naturally occult their central stars and in the process present their
vertical structure to direct view. Observations of edge-on disks require
high spatial resolution but not high contrast, so AO systems are competitive.
Furthermore, the absence of stellar PSF artifacts make edge-on systems
particularly amenable to scattered light modeling. Two of the first
known examples, HH 30 and HK Tau B, have been extensively studied over
the past few years, and are discussed at greater length in sections 2.4,
5.1, and 5.2 below. Additional examples are very valuable for
comparative studies of disk scale heights, flaring, and dust properties.
Finding new edge-on disks and imaging them across a wide range of
wavelengths are high priorities for future research.

The edge-on disk CRBR 2422.8 was discovered in the $\rho$ Oph cloud core
by {\it Brandner et al.\@} (2000a). A model for the source combining
scattered light images and mid-infrared spectra is presented by {\it
Pontoppidan et al.\@} (2005). Near $\rho$ Oph lies another new edge-on
disk, 2MASSI1628137-243139. {\it Grosso et al.\@} (2003) discovered and
present models for this object and note a peculiar near-IR color
difference between the two lobes of its reflection nebula. Three new
edge-on disks have been found via adaptive optics as companions to
brighter stars. {\it Jayawardhana et al.\@} (2002) and {\it Chauvin et
al.\@} (2002) discovered LkH$\alpha$ 263C in a quadruple system, and {\it
Jayawardhana et al.\@} present initial models suggesting a disk mass of
0.002 $M_\odot$. PDS 144 is a compact edge-on disk around a companion to
an Ae star ({\it Perrin et al.\@}, in preparation). The mysterious nature
of HV Tau C was finally resolved by {\it Monin and Bouvier} (2000) to be
an edge-on disk around the tertiary star. Modeling of HST images of HV
Tau C by {\it Stapelfeldt et al.\@} (2003) indicates that this disk also
possesses a small circumstellar envelope. Finally, a new kind of edge-on
disk has been identified around the Perseus source ASR 41: an extended
shadow from a (presumably) compact disk is cast across foreground cloud
material ({\it Hodapp et al.\@}, 2004).



\noindent{\em 2.1.3 Silhouette Disks.}
The Orion Nebula continues to be a unique and fertile ground for finding
new resolved disks. Several hundred compact ionized globules
(``proplyds'') likely contain circumstellar disks, but have a morphology
dominated by photoevaporative processing in their H II region
environment. Some of these contain internal silhouettes clearly
reminiscent of disks; and there is also a distinct category of pure
silhouette disks lacking any external ionization. In both cases, the
disk silhouettes are visible in H$\alpha$ images as foreground
absorption to the H II region. New silhouette sources are reported in a
comprehensive paper by {\it Bally et al.\@} (2000), and by {\it Smith et
al.\@} (2005). A particularly interesting new source is Orion 216-0939;
like Orion 114-426 ({\it McCaughrean et al.\@}, 1998; {\it Shuping et
al.\@}, 2003), this is a giant edge-on silhoutte more than 1000 AU in
diameter, and with bipolar reflection nebulae.

Despite the discovery of proplyds in M8 ({\it Stecklum et al.\@}, 1998)
and NGC 3603 ({\it Brandner et al.\@}, 2000b) and despite HST imaging of
NGC 2024, NGC 2264, M16, M17, and Carinae, no silhouette disks have been
identified in H II regions other than Orion. {\it M\'enard et
al.\@} (in preparation) found the first example of a silhouette disk in
nearby Taurus clouds: IRAS 04158+2805 shows a cone of scattered light, a
jet, and a silhouette 3000 AU in diameter projected in front of diffuse
H$\alpha$ emission near V892 Tau.

\subsection{Mid-Infrared Imaging}

The usual objective when imaging YSOs in the mid-infrared is to resolve
extended thermal emission from their inner disks. At the distances of
the nearest star-forming clouds, extended 10--20 $\mu$m thermal emission
has only been detected in a handful of luminous Ae stars (HD 100546, AB
Aur, and V892 Tau; {\it Liu et al.\@}, 2003, 2005; {\it Chen and Jura},
2003; {\it Pantin et al.\@}, 2005). In lower-luminosity T Tauri stars,
the 10--20 $\mu$m emssion is spatially unresolved -- even in the
relatively nearby case of TW Hya ({\it Weinberger et al.\@}, 2002). The
absence of extended 20 $\mu$m emission in T Tauri stars may pose a
problem for models of flared disk spectral energy distributions ({\it
Chiang and Goldreich}, 1997). These models postulate a superheated disk
upper layer where stellar radiation is predominantly absorbed by small
dust particles, which then radiate inefficiently at longer wavelengths.
Large grains or flatter disks may be needed to explain the fact that
most of these sources are unresolved in the mid-infrared.

One of the more significant and surprising disk imaging results of
recent years has been the detection of {\it scattered light} from three
YSO disks in the mid-infrared. {\it McCabe et al.\@} (2003) found that
the edge-on disk of HK Tau B appears as an extended source in sensitive
10 $\mu$m Keck images. The good alignment of this 10 $\mu$m nebulosity
with the optical scattered light ({\it Stapelfeldt et al.\@}, 1998), its
extent well beyond a reasonable diameter for disk thermal emission, and
its monotonically declining flux density from 2 to 10 $\mu$m argue that
this is scattered light. It is not clear if the original source of
emission is the star or the inner part of the disk. The fact that some
edge-on disks might be seen entirely via scattered light in the
mid-infrared was already indicated by {\it Infrared Space Observatory}
photometry of HH 30 ({\it Wood et al.\@}, 2002); HK Tau B is the first
resolved example. A second is in the case of GG Tau, where {\it
Duch\^ene et al.\@} (2004) clearly detect the circumbinary ring in deep
3.8 $\mu$m images. By the same arguments as above, this must also be
scattered light. A third source, PDS 144 ({\it Perrin et al.\@}, in
preparation), is an A star that appears as a spectacular bipolar nebula
in 10 $\mu$m images; the relative contributions of scattered light and
PAH emission in this image are still being assessed. As discussed in
section 5.3 below, scattered light at these long wavelengths is a
powerful diagnostic of large grains in circumstellar disks. Additional
examples of resolved mid-IR scattered light from YSO disks can be
expected in the future.

\subsection{Polarimetric Imaging}

Imaging polarimetry can confirm the presence of scattered
light nebulosity and offer clues to the location of embedded illuminating
sources.  The strength of the observed polarization depends on the
dust grain properties, the scattering geometry, the degree of multiple
scattering, and the polarization induced by any foreground
cloud material.\@  Spatially resolved polarimetry has been reported for
only a few disks.  The best example is GG Tau, where the
geometry of the circumbinary ring is very well understood,
and thus the polarimetric results can be readily interpreted.  {\it Silber
et al.\@} (2000) found that backscattered light from the far side of the
ring was very highly polarized at  1 $\mu$m, requiring that
the scattered light originate from submicron-sized grains.

A new application for polarimetry has emerged in high contrast AO.
Groundbased AO systems provide a diffraction-limited image core, but
also an extended, uncontrolled seeing halo. The light from this halo can
overwhelm the faint nebulosity of a circumstellar disk. Differential
polarimetry exploits the fact that scattered light from YSO disks is
highly polarized while the seeing halo has virtually zero polarization.
By simultaneously imaging in two polarizations, the unpolarized halo can
be removed, and the polarized disk light more clearly seen. {\it Potter}
(2003) and {\it Perrin et al.\@} (2004) report detections of several YSO
nebulosities with this technique; most are circumstellar envelopes. A
few disks have also been studied with this technique, notably TW Hya
({\it Apai et al.\@}, 2004) and LkCa 15 ({\it Potter}, 2003). The latter
has not been detected in several conventional unpolarized imaging
searches.

%

\subsection{Variability}

T Tauri stars commonly show photometric variability on timescales of a
few days to months. In the youngest stars, these variations are now
thought to be the result of hot spots, variation in accretion rate, and
occultation by warps in the inner disk, all of which are natural
consequences of magnetospheric accretion mechanism discussed in the
chapter by {\it Bouvier et al.} Disks also show variability, and in this
section we describe the best studied cases, HH~30 and AA Tau.

\noindent{\em 2.4.1 Photometric Variability in HH~30.} 
Similarly to their central stars, disks show photometric variability.
The scattered light from the disk should follow as the star brightens
and fades. The integrated magnitude of
HH 30 varies on timescales of a few days over a range of more than
one magnitude in both $V$ and $I$ ({\it Wood et al.\@}, 2000). The range is
slightly greater in $V$ than in $I$. Early photometry suggested that the
variability was periodic, but subsequent studies have not confirmed
this.

The range of variability of the disk is likely to be a lower limit on
the range of variability of the star, as the multiple optical paths
taken by scattered light will likely act to smooth the stellar
variations to some degree, both because of the finite speed of light (173
AU per day) and because the disk is illuminated by light from a range of
stellar azimuths. Cool spots produce stellar variability with ranges of
no more than one magnitude ({\it Herbst et al.\@}, 1994), and so cannot explain
the variability of the disk. Some other mechanism must be at work in HH
30, perhaps hot spots, variations in accretion rate, or occultations.
This is consistent with the strong veiling component observed by {\it White
and Hillenbrand} (2004).

\noindent{\em 2.4.2 Morphological Variability in HH 30.}
More interesting and unexpected are the quantitative changes in the
morphology of the scattered light ({\it Burrows et al.\@}, 1996; {\it
Stapelfeldt et al.\@}, 1999; {\it Cotera et al.\@}, 2001; {\it Watson and
Stapelfeldt}, in preparation). These include changes in the contrast
between the brighter and fainter nebulae over a range of more than one
magnitude; changes in the lateral contrast between the two sides of the
brighter nebula over a range of more than one magnitude (see Fig.~3);
and changes in the lateral contrast between the two sides of the fainter
nebula over a range of about half a magnitude.

\begin{figure}[tb]
 \epsscale{1.0}
 \includegraphics[width=\linewidth]{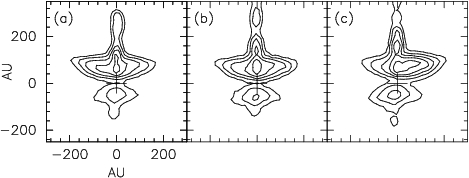} 
\caption{\small HST/WFPC2 images showing the morphological variability
of HH~30 ({\it Watson and Stapelfeldt}, in preparation). The images are
from (a) 2001 February, (b) 1995 January, and (c) 1998 March. All images
were taken in a broad filter centered at 675~nm. The images show the
left side brighter, both sides having similar brighnesses, and the right
side brighter.}
\end{figure}

The timescales for the variability are uncertain, but are less than one
year. Thus, the variability is not due to changes in the disk structure
at radii of 100 AU but rather changes in the pattern of illumination at
radii of 1 AU or less. Thus, this \emph{morphological} variability is
fascinating as it allows us to peek into the central part of the disk
and possibly constrain the geometry of the accretion region.

However, before we use the variability to constrain anything, we need to
understand its origin. Two mechanisms have been suggested; {\it Wood and
Whitney} (1998) have suggested that inclined hot spots could illuminate
the outer parts of disk like lighthouses, and {\it Stapelfeldt et al.\@}
(1999) have suggested that warps in the inner disk could cast shadows
over the outer disk.

In the case of HH 30, one might hope to distinguish between the
mechanisms on the basis of the color of the morphological variability.
HST observations show no significant differences between $V$ and $I$.
Naively, one would expect illumination by hot spots to produce stronger
variability in the blue whereas optically-thick shadowing should be
neutral in color. However, spectroscopic studies show that the veiling
component is often relatively flat in the red ({\it Basri and Batalha},
1990; {\it White and Hillenbrand}, 2004), so the lack of color is
inconclusive.

\begin{deluxetable}{ll}
\tablecolumns{1}
\tablewidth{0pt}
\tabletypesize{\small}
\tablecaption{Publicly Available Scattering and Thermal Equilibrium Codes}
\tablehead{Code&URL}
\startdata
HO-CHUNK&http://gemelli.colorado.edu/\~\space bwhitney/codes/codes.html\\
MC3D&http://www.mpia-hd.mpg.de/FRINGE/SOFTWARE/mc3d/mc3d.html\\
Pinball&http://www.astrosmo.unam.mx/\~\space a.watson/\\
RADMC and RADICAL&http://www.mpia-hd.mpg.de/homes/dullemon/radtrans/\\
\enddata
\end{deluxetable}  

Determining the timescale might help to distinguish between the
mechanisms, although if the putative warp in the inner disk is locked to
the star as it appears to be in AA Tau (discussed below) this will also
be inconclusive. HST observations show that the asymmetry can change
from one side to the other in no more than about six months but do not
significantly constrain a lower limit on the period or, indeed,
demonstrate that the variability is periodic rather than stochastic.
Since the net polarization vectors of the two sides of the bright nebula
are not parallel (see Fig.~6 of {\it Whitney and Hartmann}, 1992), the
asymmetry should produce a polarimetric signature. On-going polarimetric
monitoring of HH~30 may help to determine the timescale of the
asymmetry.

\noindent{\em 2.4.3 AA Tau.}
AA Tau seems to be a prototype for both mechanisms suggested to explain
the morphological variability of HH 30, apparently possessing both
inclined hot spots and occulting inner-disk warps ({\it Bouvier et al.\@},
1999; {\it M\'enard et al.\@}, 2003; {\it O'Sulivan et al.\@}, 2005), both
the result of an inclined stellar magnetic dipole. The disk around AA
Tau has recently been imaged in scattered light by {\it Grady et al.\@}
(in preparation), although previous observations with similar
sensitivities did not detect the disk. It seems that the disk was
finally detected because it was observed at an epoch in which the
inner-disk warp was at least partially occulting the star.

\noindent{\em 2.4.4 Other Disks.}
It would be useful to be able to
compare the variability in HH~30 to other disks, to understand what is
unusual and what is common. In this respect, recent second-epoch
observations of several objects by {\it Cotera et al}.\ (in preparation)
are very useful. Nevertheless, since most disks have been observed at
only one or two epochs, there is little that we can say about their
variability.

\section{MODELING TECHNIQUES}


\subsection{Radiation Transfer}


There are several techniques for simulating scattered light images of
disks and envelopes including the single scattering approximation (e.g.,
{\it Dent}, 1988; {\it Burrows et al.\@}, 1996; {\it D'Alessio et al.\@},
1999) and direct integration of the equation of radiation transfer under
the assumption of isotropic scattering (e.g., {\it Dullemond and
Dominik}, 2004). However, by far the most common techniques are Monte
Carlo simulations or integrations (e.g., {\it Lef\`evre et al}, 1982,
1983; {\it Bastien and M\'enard}, 1988, 1990; {\it Whitney and
Hartmann}, 1992, 1993; {\it Lopez, M\'ekarnia, and Lef\`evre}, 1995; {\it
Burrows et al.\@}, 1996; {\it Whitney et al.\@}, 1997, 2003a,b, 2004; {\it
Lucas and Roche}, 1997, 1998; {\it Stapelfeldt et al.\@}, 1999; {\it Wood
et al.\@}, 1998, 1999, 2001; {\it Lucy}, 1999; {\it Wolf et al.\@}, 1999,
2002, 2003; {\it Bjorkman and Wood}, 2001, {\it Cotera et al.\@}, 2001;
{\it Watson and Henney}, 2001; {\it Schneider et al.\@}, 2003; {\it
Stamatellos and Whitworth}, 2003, 2005; {\it Watson and Stapelfeldt},
2004). Faster computers and improved algorithms allow these simulations
to be fast, incorporate anisotropic scattering, polarization, and fully
three-dimensional circumstellar geometries and illuminations.

For optical and near-infrared simulations one can normally assume that
scattered {\it starlight} dominates the images and there is no
contribution from dust reprocessing. With this assumption, an image at a
specific wavelength and orientation can be calculated in a few minutes
on current computers. Simulations at longer wavelengths must include
thermal reprocessing and calculate thermal equilibrium (e.g., {\it
D'Alessio et al.\@}, 1998, 1999; {\it Whitney et al.\@}, 2003a,b).
                                     
Some groups have made their scattering and
thermal equilibrium codes publicly available (see Table 2), and these
tools are now being used by the community.

\subsection{Density Distributions}

Modelers take different approaches to the density distribution in
disks. One extreme is to assume disks are vertically isothermal, the dust is
well mixed with the gas, and the surface density and scale heights are
power laws in the radius. Another extreme is to solve self-consistently
for the temperature using thermal equilibrium, to solve for the vertical
density distribution using pressure equilibrium, to solve for the
surface density assuming an accretion mechanism with a constant
mass-transfer rate, and to include dust settling. In between are many
intermediate approaches, for example, solving for thermal equilibrium
and vertical pressure equilibrium but imposing a surface density law.

\begin{deluxetable}{ll}
\tablecolumns{1}
\tablewidth{0pt}
\tabletypesize{\small}
\tablecaption{Parameters Constrained by Scattered-Light Images}
\tablehead{Inclination&Parameters}
\startdata
Edge-on&inclination, mass-opacity, forward scattering, and scale height\\
Intermediate&inclination, mass-opacity, forward scattering, and outer radius\\
Face-on&inclination, radial dependence of scale height\\
\enddata
\end{deluxetable}  

The approaches are complementary. The simple power-law disks have very
little physics, but have many ``knobs'' that can be arbitrarily adjusted
to represent a wide range of disk density distributions and thereby
cover the very real uncertainties in our understanding of these objects.
On the other hand, the approaches that incorporate more and more physics
have fewer and fewer ``knobs''. In one sense, they are more realistic,
but only to the extent that our understanding of disk physics is
correct. Unfortunately, there are real gaps in our knowledge. For
example, thermal equilibrium depends on the dust opacity, which is not
well known; the details of accretion are still actively being
researched, with disk viscosities uncertain by orders of magnitude; and
disks may not have a constant inward mass-transfer rate.

Both approaches are useful, but is important to understand the strengths
and weaknesses of each. Simple parameterized models allow one to
investigate the dependence of other properties on the density
distribution (e.g., {\it Chauvin et al.\@}, 2002; {\it Watson and
Stapelfeldt}, 2004), but are limited in what they can tell us about disk
physics. On the other hand, models with more physics are necessary to
test and advance our understanding of disks (e.g., {\it D'Alessio et
al.\@}, 1998, 1999; {\it Schneider et al.\@}, 2003; {\it Calvet et al.\@},
2005), but our knowledge of the input physics in these models is still
incomplete. For these reasons, future modeling efforts will continue to
tailor their approaches to modeling the density distribution according
to the specific problem being addressed.

\subsection{Fitting}

Determining how well scattered light images can be modeled is important
for testing physical models of flared disks and collapsing envelopes.
Many studies have been succesful at reproducing the overall morphology
and intensity pattern of scattered light images (e.g., {\it Lucas and
Roche}, 1997; {\it Whitney et al.\@}, 1997; {\it Wood et al.\@}, 2001).
However, quantitative model fitting is more convincing and provides a
better test of physical models. {\it Burrows et al.\@} (1996) and {\it
Krist et al.\@} (2002) applied least-squares fitting techniques to
single-scattering models of the disks of HH~30 and HK Tau B and
determined their physical structure and scattering properties. More
recently, {\it Watson and Stapelfeldt} (2004) applied least-squares
fitting techniques to multiple-scattering models of the disk of HH~30
and allowed the density structure and dust scattering properties to be
free parameters. Their results provide strong constraints on the
circumstellar density, structure, and dust properties (see section 5.2).
{\it Glauser et al.\@} (in preparation) have recently fitted models to
scattered-light images, polarization images, and the SED of IRAS
04158+2805.

%





\section{DISK STRUCTURE FROM SCATTERED LIGHT IMAGES}

Interpreting scattered light images of optically-thick disks is an
inverse problem. The physics that takes us from a three-dimensional
distribution of emissivity and opacity to a two-dimensional distribution
of surface brightness removes a great deal of information. In this
section we discuss what information is lost and what can be recovered.
The orientation from which a disk is viewed is a critical factor in
this. For this reason, we classify disks as edge-on,
intermediate-inclination, and face-on. The parameters that can be most
directly constrained in each case are summarized in Table~3.

\subsection{Edge-On Disks}

\begin{figure}[t]
\epsscale{1.0}
 \includegraphics[width=\linewidth]{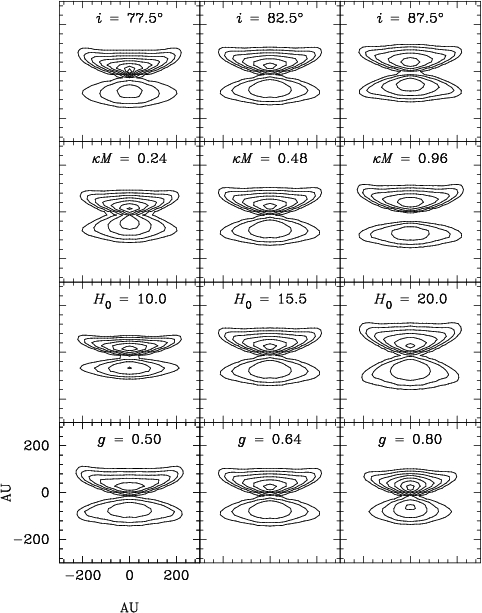}
 \caption{\small Scattered-light models of an edge-on disk. The center
 column shows model A1 of {\it Burrows et al.\@} (1996). The left and
 right columns show models with, from top to bottom, different
 inclinations from face-on, different mass-opacity products $\kappa M$
 (in $\mathrm{cm^2\,g^{-1}\,M_\odot}$, scale height normalizations $H_0$
 (in AU), and phase function asymmetry parameters $g$. The contours have
 the same level in each panel and are spaced by factors of 2. See {\it
 Burrows et al.\@} for precise definitions of the density distribution and
 parameters.}
\end{figure}

Opaque material close to the equatorial plane of edge-on optically-thick
disks occults the star; all that is seen at optical and near-infrared
wavelengths are two nebulae formed by light scattered by material away
from the equatorial plane. These nebulae tend to be dominated by
material in the segment of the disk closest to the observer. The
information present in the images can be summarized as follows:

\begin{enumerate}

\item The brightness ratio of the nebulae. This is largely sensitive to
  the inclination of the disk with respect to the observer. When a
  symmetrical star-disk system is observed in its equatorial plane, the
  nebulae should have equal brightness. As disk is tilted with respect
  to the observer, the nebula that is closer to the line of sight to the
  star dominates (see the top row of Fig.\ 4).

\item The minimum separation of the nebulae. The minimum separation
  occurs on the projected axis of an axisymmetric disk. It is largely
  sensitive to the total mass-opacity product of the disk. As the
  mass-opacity product of a disk is increased, the dark lane separating
  the two nebulae widens (see the second row of Fig.\ 4). A good example
  of this is the decreasing separation of the nebulae of HH 30 with
  increasing wavelength (e.g., Figs.~6 and 8 of {\it Cotera et al.\@},
  2001, and Fig.~2 of {\it Watson and Stapelfeldt}, 2004). One cannot
  use the width of the dark lane to constrain the mass or the absolute
  opacity individually without additional information. This is because
  the appearance of the disk at optical and near-infrared wavelengths is
  a scattering problem, and in the equations governing such problems the
  mass density and opacity per unit mass always appear as a product. On
  the other hand, one can use changes in the width of the lane to
  quantify relative changes in the opacity as a function of wavelength
  (see section 5.2).


\item The increase in separation of the nebulae with increasing
  projected distance from the star. By this we refer to the degree of
  curvature of the boundary of the dark lane separating the two nebulae
  or the degree of apparent flaring of the nebulae. This is largely
  sensitive to the effective scale height of dust in the outer part of
  the disk (say, at the radii which dominate the observed scattered
  light). Larger scale heights lead to greater curvature or flaring (see
  the third row of Fig.\ 4). Good examples of this are the relatively
  flared nebulae of HH 30, which has a scale height of 16--18 AU at 100
  AU, and the very flat nebulae of HK Tau B, which has a scale height of
  about 4 AU at 50 AU (see section 5.1).


\item The vertical extent of each individual nebula. By this we mean the
  apparent height of each nebula in comparison to its apparent diameter.
  Again, this is largely sensitive to the effective scale height of dust
  in the outer part of the disk. Larger scale heights lead to more
  extended nebulae (see the third row of Fig.\ 4). Good examples of this
  are again the vertically extended nebulae of HH 30 and the narrow
  nebulae of HK Tau B.

\item The degree to which the surface brightness of the nebulae drops
  with increasing projected distance from the star. This is largely
  sensitive to the degree to which forward scattering dominates the
  phase function. Enhanced forward scattering produces a more
  centrally-concentated distribution and a more pronounced decrease in
  surface brightness with increasing projected distance (see the fourth
  row of Fig.\ 4). The light that emerges close to the projected disk
  axis has scattered through relatively small angles and the light that
  emerges far from the projected disk axis has scattered through
  relatively large angles. Increased forward scattering enhances the
  former and diminishes the latter. A good example of this is the better
  fits for HH 30 obtained by {\it Watson and Stapelfeldt}\ (2004) with
  high values of the phase function asymmetry parameter $g$.

\end{enumerate}

However, much information that is lost or ambiguous:

\begin{enumerate}

\item The luminosity of the star. All of the light we see is processed
  through the disk, either by scattering or by absorption and thermal
  emission.

\item The inner region of the disk. The observed disks appear to be
  sufficiently flared that the light scattered from the outer parts
  comes from a relatively small range of angles and passes either
  completely above the inner disk or through regions of the inner disk
  with similar extinctions. In these case, one could remove the inner 30
  AU of the disk and the observed pattern of scattered light would not
  change, although the total brightness of scattered light might.

  An exception to this general rule may be HH 30, in which is it is
  possible that parts of the inner disk shadows parts of the outer disk
  (see section 2.4). Another exception may be disks in which the inner
  part of the disk shadows the outer part completely.

\item The outer radius. Scattered light images of edge-on disks
  sometimes show a relatively sharp outer cut-off and this is often
  interpreted as the outer radius of the disk. However, it is possible
  that the disk continues beyond the bright nebulae, but that it is
  shadowed. Such an extension would likely not be seen in scattered
  light. There is evidence that this is the case in the sihouette disk
  Orion 114-426, where the scattered light nebulae do not extend to the
  edge of the sihouette (see Fig.~2d of {\it McCaughrean et al.\@},
  1998). An extended disk like this might also be detectable in CO.

\item The radial dependence of the surface density and effective scale
  height. Neither of these have strong effects on the models, at least
  as long as they are constrained to lie within plausible ranges. Worse,
  even their subtle effects are degenerate; putting more mass at larger
  radii produces similar effects in the images to increasing the scale
  height at larger radii ({\it Burrows et al.\@}, 1996; {\it Watson and
  Stapelfeldt}, 2004).

\end{enumerate}

It is clear from this that scattered light images of edge-on
optically-thick disks provide limited but important information on disk
properties. Those which can be most cleanly separated are the
inclination, the relative opacities at different wavelengths, the scale
height in the outer part of the disk, and the degree of forward
scattering in the phase function.

\subsection{Intermediate-Inclination Disks}

\begin{figure}[tb]
\epsscale{1.0}
 \includegraphics[width=\linewidth]{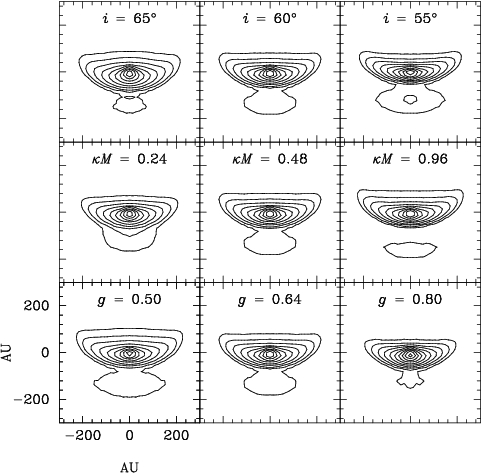}
 \caption{\small Scattered-light models of an intermediate-inclination
 disk. The center column shows model A1 of {\it Burrows et al.\@} (1996)
 inclined to 60 degrees from face-on. The left and right columns show
 models with, from top to bottom, different inclinations from face-on,
 different mass-opacity products $\kappa M$ (in
 $\mathrm{cm^2\,g^{-1}\,M_\odot}$, and phase function asymmetry
 parameters $g$. The contours have the same level in each panel and are
 spaced by factors of 2. See {\it Burrows et al.\@} for precise
 definitions of the density distribution and parameters.}
\end{figure}

Intermediate-inclination disks are those in which the star is directly
visible but both nebulae are still present (although one may be lost in
the noise). In contrast to edge-on disks, in principle one nebulae
extends over all azimuths from the star, although some azimuths may be
lost in the glare of direct light from the star. Scattered-light images
of intermediate-angle optically-thick disks have not been as extensively
studied as edge-on disks, but preliminary studies ({\it Schneider et
al.\@}, 2003; {\it Quijano}, 2005) suggest that the following information
is present:

\begin{enumerate}

\item The brightness ratio of the nebulae. This depends principally on
  the mass-opacity product and the inclination (see the top two rows of
  Fig.\ 5). 


\item The separation of the nebulae. Like the brightness ratio, this
  depends principally on the mass-opacity product and the inclination
  (see the top two rows of Fig.\ 5).

\item The degree to which the surface brightness drops with increasing
  projected distance from the star. This depends principally on the the
  degree to which forward scattering dominates the phase function (see
  the third row of Fig.\ 5).


\item The ratio of scattered light to unscattered light. That is, the
  relative brightness of the nebulae and the star. This depends on the
  scale height in the outer disk, the albedo, the degree to which
  forward scattering dominates the phase function, the inclination, and
  the mass-opacity product. (A complication here is that light scattered
  very close to the star is difficult to distinguish from direct light
  from the star.)

\item The outer radius. The presense of two nebulae in many systems
  suggests that there is not a cold, collapsed, optically-thick disk
  that extends significantly beyond the bright nebulae. This allows one
  to constrain the outer radius of the optically-thin disk with some
  degree of confidence. For example, GM Aur shows two nebulae which
  suggest that the true outer radius of the optically-thick disk is
  around 300 AU ({\it Schneider et al.\@}, 2003).



\end{enumerate}

Again, much information is lost:

\begin{enumerate}

\item The inner region of the disk. In current images, information on
  the inner disk is lost under the PSF of the star, even when observing
  in polarized light (e.g., {\it Apai et al.\@}, 2004). Infrared 
  interferometry, covered in the chapter by {\it Millan-Gabet et al.\@},
  can provide a great deal of information on the very innermost part of
  the disk at radii of less than 1 AU. Coronagraphic images from space
  or, perhaps, from extreme AO systems will be required to recover
  information on scales larger than those available to interferometers
  but smaller than those lost under current PSFs.

\item The radial dependence of the surface density and effective scale
  height. As with edge-on disks, neither of these have strong effects on
  the observed morphology.

\end{enumerate}

Scattered-light images of intermediate-angle optically-thick disks are
thus more difficult to interpret than those of edge-on disks. The only
parameter that seems to be cleanly separated is the outer radius of the
disk. The mass-opacity product and the inclination are to a large degree
degenerate (compare, for example, the first two rows of Fig.\ 5, which
show similar changes in the brighter nebula), but if millimeter data or
an SED are available to constrain the inclination, the mass opacity
product can be obtained from scattered-light images. The dust asymmetry
parameter can be constrained if disk structure is known with adequate
certainty.

\subsection{Face-On Disks}

\begin{figure}[tb]
\epsscale{1.0}
 \includegraphics[width=\linewidth]{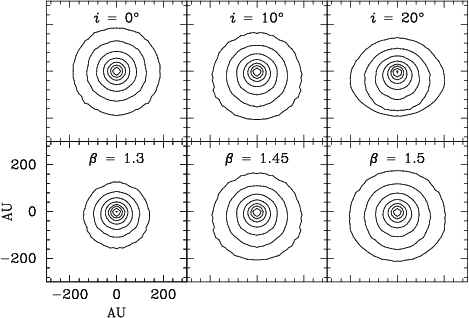}
 \caption{\small Scattered-light models of a face-on disk. The center
 column shows model A1 of {\it Burrows et al.\@} (1996) inclined to 10
 degrees from face-on. The left and right columns show models with, from
 top to bottom, different inclinations from face-on and different
 indicies $\beta$ in the scale-height power law. The contours have the
 same level in each panel and are spaced by factors of 2. See {\it
 Burrows et al.\@} for precise definitions of the density distribution and
 parameters.}
\end{figure}

Face-on disks are those in which departures from radial symmetry are
dominated by inclination (combined with the phase function) rather than
vertical structure (at least for axisymmetric star-disk systems). The
only example we have of such a disk around a YSO is TW Hya ({\it Krist
et al.\@}, 2000; {\it Trilling et al.\@}, 2001; {\it Weinberger et al.\@},
2002; {\it Apai et al.\@}, 2004; {\it Roberge et al.\@}, 2005).

The information present in scattered-light images of face-on
optically-thick disks is:

\begin{enumerate}

\item The ellipticity. This is by definition determined by the
  inclination and phase function in axisymmetric star-disk systems.
  Departures from constant ellipticity can provide a fascinating insight
  into non-axisymmetric illumination or disk structure. For example,
  {\it Roberge et al.\@} (2005) show evidence that the disk around TW Hya
  is significantly elliptical between 65 and 140 AU and essentially
  circular beyond this region; they suggest that this suggests that the
  disk is warped. (See the first row of Fig.\ 6.)

\item The dependence of the surface brightness on radius. This depends
  essentially on the radial dependence of the scale height. If the disk
  is well-mixed, this provides information on the radial dependence of
  the temperature. (See the second row of Fig.\ 6.)

\item The ratio of scattered light to unscattered light. That is, the
  relative brightness of the nebula and the star. This depends
  principally on the scale height in the outer disk, the albedo, the
  phase function, and the mass-opacity product.

\end{enumerate}

Information on the inner disk is not present for the same reasons as in
the case of intermediate-inclination disks. Information on the outer
disk radius is not present for the same reasons as in the case of
edge-on disks.

\section{HIGHLIGHTS OF APPLIED DISK MODELING}

In this section, we present four examples of scientific results from the
modeling of scattered-light disks around YSOs. Our first three are
related to the evolution of dust, which is also discussed in the
chapters by {\it Natta et al.\@} and {\it Dominik et al.} The fourth
concerns evidence for an inner hole in the disk around GM Aur, which may
have been cleared by a planet (see the chapter by {\it Papaloizou et al.\@}).
While many disks have been imaged, the cases discussed below are perhaps
the best examples of how disk structure and dust properties have been
derived from modeling of multi-wavelength image datasets.

\subsection{Evidence for Dust Settling}

\begin{deluxetable}{lrrl}
\tablecolumns{1}
\tablewidth{0pt}
\tabletypesize{\small}
\tablecaption{Scale Heights}
\tablehead{Object & $H$ at 50 AU &$T_\mathrm{equivalent}$&Reference}
\startdata
HH 30        & 6.3 AU & 51 K &{\it Watson and Stapelfeldt} (2004)\\
IRAS 04302+2247   & 6.1 AU & 48 K&{\it Wolf et al.\@} (2003) \\
HV Tau C     & 6.5 AU & 35 K &{\it Stapelfeldt et al.\@} (2003)\\
GM Aur       & 3.4 AU & 18 K &{\it Schneider et al.\@} (2003)\\
HK Tau B     & 3.8 AU &  8 K &{\it Stapelfeldt et al.\@} (1998)\\
\enddata
\end{deluxetable}  

The gas scale height in disks is determined by the balance between the
stellar gravitational force which compresses the disk toward its
midplane and gas pressure which acts to puff the disk up. In vertical
hydrostatic equilibrium in a vertically isothermal disk, the gas scale
height is related to the local {\it gas temperature} according to $H(r)=
\sqrt{kT(r)r^3 / G M_{*}m}$, where $m$ is the mean molecular weight of
the disk gas. Direct determination of the disk temperature is difficult,
but reasonable results can be obtained by modeling the infrared and
millimeter spectrum. A complication is that disks likely have vertical
temperature inversions ({\it Calvet et al.\@}, 1991, 1992; {\it Chiang and
Goldreich}, 1997; {\it D'Alessio et al.\@}, 1998).

If the dust and gas are well mixed, then the dust follows the gas
density distribution and will have the same scale height. As we have
seen in section 4.1, the dust scale height can be determined from
high-resolution scattered-light images. It is important to note that the
height of the scattering surface above the disk plane is not equal to
the dust scale height; rather, the scattering surface defines the locus
of points with optical depth unity between the star and individual disk
volume elements, and can be any number of dust scale heights above the
midplane. The primary observational indicators of the scale height are
the vertical extent of the nebula, the degree of curvature of the
nebulae, and the sharpness of the dark lane. Disks with large scale
heights appear more vertically extended (``fluffy'') and have curved
disks whereas those with small scale heights are vertically narrow and
have almost parallel nebulae. Exact value for the scale height must be
determined by fitting models to the images.

The comparison of the scale heights for the gas and dust offer a unique
opportunity to test the assumption that disks are vertically well mixed.
{\it Burrows et al.\@} (1996) derived an equivalent temperature from the
dust scale height for the HH 30 disk at 100 AU radius under the
assumption that the disk was vertically well mixed. The result was
broadly consistent with expectations from a simple thermal model, and
suggested that the assumption that the disk was vertically well mixed
was correct.

Since {\it Protostar and Planets IV} dust scale heights have been
derived for several other YSO disks. In two cases, HV Tau C, and IRAS
04302+2247, initial scale height derivations implied unreasonably high
equivalent temperatures and it is difficult to imagine why the dust
would be more extended than the gas ({\it Stapelfeldt et al.\@}, 2003;
{\it Wolf et al.\@}, 2003). These appear to be caused by the presence of
circumstellar envelopes in addition to the disks, with the envelopes
producing more diffuse nebulosity than expected for a pure disk. By
adding an envelope to the density distribution, it was possible to
remove this effect, and derive dust scale heights more consistent with
simple thermal models. Scale height values for these and three other
disks are shown in Table~4. The scale heights in the original references
have been extrapolated to a reference radius of 50 AU to facilitate
comparison.

The dust scale heights measured to date fall into two groups with the
values differing by almost a factor of two. The three objects with
larger scale heights all have outflows indicating ongoing accretion,
whereas the two objects with smaller scale heights have little or no
outflow activity. This could be observational evidence that accreting
disks are systematically more puffed-up (warmer) than non-accreting
disks. The small equivalent temperature implied for the HK Tau B disk
merits particular attention. Unfortunately the infrared spectral energy
distribution for this disk is incomplete, so this value cannot be
compared to a well-constrained disk thermal model. However, it is highly
unlikely that the gas in the disk could actually be as cold as 8 K.
Instead, this could be a case where the assumption that the dust and gas
are well mixed is not correct. Instead, it appears more likely that the
dust has decoupled from the gas and partially settled toward the disk
midplane, an expected stage of disk evolution ({\it Dubrulle et al.\@}
1995; {\it Dullemond and Dominik}, 2004). In this case, the equivalent
temperature derived from the dust scale height would be lower than the
true gas temperature.

The vertical scale height of dust in a YSO circumstellar disk could be
a key indicator of its structure and evolutionary state. It would be
very valuable to accumulate scale height measurements for larger numbers
of edge-on disks, to see how unique the HK Tau B results are and uncover
the full diversity of scale heights in the YSO disk population.

\subsection{Evidence for Dust Growth from the Opacity Law}

In optically thin media, the color of scattered light directly depends
on the wavelength dependence of the grain opacity. Small grains like
those in the ISM have a much higher scattering cross section at shorter
wavelengths, and thus optically thin nebulae will have strongly blue
colors relative to their illuminating star. However, YSO disks are
optically thick in the optical and near-infrared. In this situation the
color of the scattered light no longer depends on the grain opacity, and
the disk will appear spectrally neutral relative to the star -- even if
small grains are the dominant scatterers. Neutral colors have been
observed for the disks of TW Hya ({\it Krist et al.\@}, 2000; {\it
Weinberger et al.\@}, 2002; {\it Roberge et al.\@}, 2005) and GM Aur ({\it
Schneider et al.\@}, 2003), consistent with this expectation. A mechanism
that can produce non-stellar colors in reflection from an optically
thick disk is a wavelength-dependent dust albedo or phase function, but
this is thought to be a small effect. Given these considerations, how
can scattered light from an optically thick disk be used to constrain
the grain properties?

The answer is to study changes in nebular spatial structure as a
function of wavelength. For small ISM-like particles, the $\tau=1$
scattering surface is located in lower-density regions above the disk
midplane at optical wavelengths, and shifts into higher-density regions
nearer the midplane at near-infrared wavelengths. Conversely, large
grains acting as grey scatterers would produce a reflection nebula whose
spatial structure would not vary with wavelength. 

\begin{figure}[tb]
\epsscale{1}
 \includegraphics[width=\linewidth]{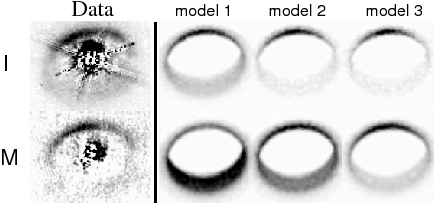}
 \caption{\small A comparison of data and models for the GG Tau
   circumbinary ring, in the $I$ band (0.8 $\mu$m; top row) and $M$ band
   (3.8 $\mu$m; bottom row). The ring is inclinced 37$^{\circ}$ from
   face on, with the forward scattering edge projected above the binary,
   and backscattering side projected below the binary. The key
   observable is the brightness contrast between the front and back
   sides. The first model has a maximum grain size of 0.3 $\mu$m. It
   matches the $I$ band image well, but in the $M$ band it predicts too
   much backscattering and not enough forward scattering. The second
   model has a maximum grain size of 0.9 $\mu$m. In this model, the
   backscattered flux becomes too small in the $I$ band, and still too
   large in the $M$ band. A reasonable mach to the $M$ band image is
   obtained in the third model (maximum grain size of 1.5 $\mu$m), but
   the $I$ band backscattering is underpredicted. No single grain size
   distribution accounts for the phase function effects at both
   wavelengths. Figure and results from {\it Duch\^ene et al.\@} (2004).}
\end{figure}

For edge-on disks, the key observable is a narrowing of the central dust
lane as the object is imaged at progressively longer wavelengths (see
the second row of Fig.~4). This behavior is seen in Orion 114-426 ({\it
McCaughrean et al.\@}, 1998), IRAS 04302+2247 ({\it Padgett et al.\@},
1999), HH 30 ({\it Cotera et al.\@}, 2001), and HV Tau C ({\it Stapelfeldt
et al.\@}, 2003). It provides clear evidence that the scattering dust
grains are dominated by small particles. But how small? A quantitative
answer can be derived by fitting scattered light models to
multiwavelength image data sets. This was first done in the case of HH
30 by {\it Cotera et al.\@} (2001). From modeling of HST near-infrared
images, they found that the dust lane thickness changed less quickly
with wavelength than expected for standard interstellar grains, and
interpreted this as evidence for grain growth in the disk. {\it Watson
and Stapelfeldt} (2004) extended this analysis by including optical
images and by considering a wider range of possible density structures,
but found essentially the same result: the most likely ratio of grain
opacities between 0.45 and 2.0 $\mu$m is 2.0 for the HH 30 disk, versus
a value of 10 expected for ISM grain models.

While the disk of HH 30 appears to show some grain evolution, modeling
of multiwavelength images of the edge-on disks of HV Tau C ({\it
Stapelfeldt et al.\@}, 2003) and IRAS 04302+2247 ({\it Wolf et al.\@}, 2003)
finds grain opacity ratios consistent with standard ISM grains. Both of
these sources possess circumstellar envelopes in addition to disks, so
the presence of primitive grains could reflect this ongoing infall from
the ISM onto the disks. The disk of HK Tau B shows only subtle changes
in its dust lane thickness between optical and near-IR images, and may
represent a more evolved system; a firm conclusion on its dust
properties can be expected from future model fitting. Dust properties in
the giant edge-on silhouette disk Orion 114-426 are uncertain; {\it
Throop et al.\@} (2001) found the radial extent of the silhouette to be
achromatic between 0.66 and 1.87 $\mu$m; {\it Shuping et al.\@} (2003)
found that it was chromatic between 1.87 and 4.05 $\mu$m; while {\it
McCaughrean et al.\@} (1998) showed that the dust lane thickness between
the lobes of reflected light was clearly chromatic between 1.1 and 2.0
$\mu$m. Additional modeling of this source is needed. Observations and
modeling of a broader sample of edge-on disks offer an opportunity to
probe the diversity of dust properties across the variables of disk age,
disk enviroment, and accretion signatures, and should be vigorously
pursued.

%
%
%
%
%
%
%
%
%



\subsection{Evidence for Dust Stratification from the
 Phase Function}

GG Tau is a binary T Tauri star with 0.3\asec (42 AU) projected
separation. It hosts the most prominent example of a circumbinary ring
of dust and gas. The ring has been studied in the millimeter lines and
continuum ({\it Guilloteau et al.\@}, 1999; {\it Wood et al.\@}, 1999),
near-infrared scattered light ({\it Roddier et al.\@}, 1996; {\it Wood et
al.\@}, 1999; {\it Silber et al.\@}, 2000; {\it McCabe et al.\@}, 2002), and
optical scattered light ({\it Krist et al.\@}, 2002, 2005). Through these
studies, the density structure of the GG Tau ring is perhaps now the
best understood of all YSO disks. A key feature of the ring is its
intermediate inclination of 37$^{\circ}$ from face on. This spatially 
separates both the foreground and background parts of the ring from each 
other and from the central binary. This ``clean'' configuration allows 
the relative strength of forward scattering and backscattering to be directly
measured. This quantity can be a powerful diagnostic of dust properties
in the ring.

The scattering phase function strongly favors forward scattering when
the grain size is comparable to the wavelength. At
wavelengths much larger than the grain size, scattering becomes
more isotropic. A comprehensive study of
phase function effects in the GG Tau ring was recently carried out by
{\it Duch\^ene et al.\@} (2004). Using new images taken at 3.8
$\mu$m, the longest wavelength to date at which scattered light from the
ring has been detected, and existing images at shorter
wavelengths, {\it Duch\^ene et al.\@} modeled the wavelength dependence of
the phase function from 0.8 to 3.8$\mu$m. Highlights from their
results are shown in Fig.~7. The results show that the 3.8 $\mu$m
scattered light must arise from dust grains larger than those in the
ISM, whereas the 0.8 $\mu$m scattered light must simultaneously
originate from much smaller particles. Additional evidence for small
particles is provided by the 1.0 $\mu$m polarimetry of {\it Silber et
al.\@} (2000), who found the backscattered light was highly polarized and
thus dominated by submicron grains.

The GG Tau phase function results indicate that no single power-law
distribution of grain sizes can simultaneously account for the
observations. {\it Duch\^ene et al.\@} (2004) suggest that these results
can be explained by a vertically stratified disk in which the grain size
increases towards the midplane. In this view, large grains responsible
for the 3.8 $\mu$m scattered light would be located in a denser region
closer to the disk midplane, a region that shorter wavelength photons
cannot reach. The stratification could be due to dust settling to the
ring midplane. Alternatively, it might be the case that the preference
for scattering at grain sizes comparable to the wavelength is so strong
that the large grains dominate the 3.8 $\mu$m scattered light, even
though they are less numerous than the small grains. In that case,
vertical dust settling would not necessarily have taken place. 

This picture is very attractive. However, {\it Krist et al.\@} (2005) point out
that current models of GG Tau do not fully explain the observations. For
example, no current model simultaneously reproduces the total brightness
of the disk, its color, and its azimuthal variation. Furthermore, the
ratio of brightness of the near and far parts of the disk has been
observed to vary with time. Additional observations and models will be
required to confirm the suggestion of a stratified disk. It would be
very valuable to perform a similar study in other disk systems;
unfortunately, the GG Tau ring is currently unique.

\subsection{Combining Scattered Light Images and SEDs}

GM Aur has a disk viewed at an intermediate inclination. The disk is
clearly seen in WFPC2 and NICMOS images after careful subtraction of
a reference PSF. Scattered light is detected between 0.4{\asec} and
2.1\asec (55 and 300 AU) from the star. Modeling of the scattered light
images by {\it Schneider et al.\@} (2003) derived a dust vertical scale height
of 8 AU at a radius of 100 AU and a disk inclination 56$^\circ$ from
face-on, and demonstrated that there was also a remnant circumstellar
envelope. However, the coronagraphic occulting spot used for the
observations blocked any information on the properties of the inner
disk. To access that region, {\it Schneider et al.\@} turned to the infrared
spectral energy distribution (SED). Emission between 2 and 70 $\mu$m probes
the disk temperature structure inside 40 AU, and thus is complementary
to the results of the scattered light modeling. Using the disk model
derived from the scattered light images, and appropriate assumptions
about grain properties, {\it Schneider et al.\@} were able to reproduce GM Aur's
SED, including the millimeter continuum points (see Fig.~8). The very
small near-IR excess emission requires that the inner part of the disk
be optically thin. {\it Rice et al.\@} (2003) showed that the inner 
region may have been cleared by a Jovian-mass planet.

More recently, {\it Calvet et al.\@} (2005) have presented {\it Spitzer} IRS
spectroscopy of GM~Aur. This data requires an inner optically-thin
region that extends to 24~AU from the star. This is much larger than
previous estimates based on modeling of broad-band mid-infrared
photometry. The {\it Spitzer} IRS data show that the inner disk is not
empty, but contains a small amount of small dust grains which produce
the silicate emission feature. To fit the long wavelength SED, the
average grain size in the outer disk must be larger than in the inner,
optically-thin region. 

\begin{figure}[tb]
\epsscale{0.9}
 \includegraphics[width=\linewidth]{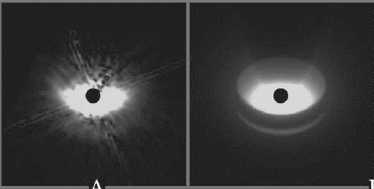}
\bigskip
\epsscale{1.0}
 \includegraphics[width=\linewidth]{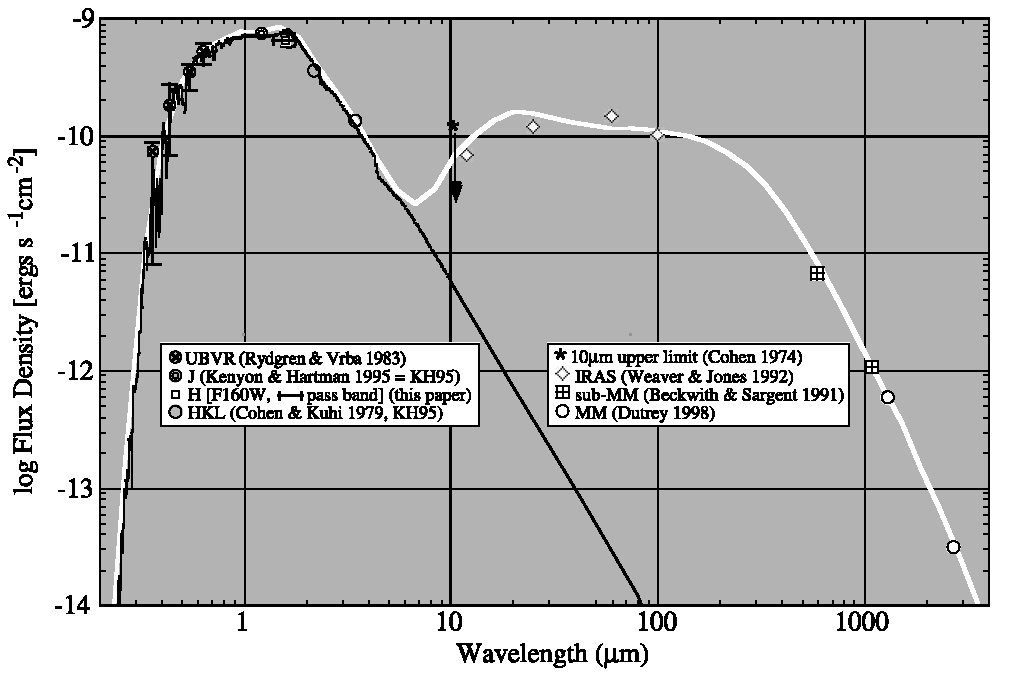}
 \caption{\small Scattered light images and SEDs of the classical T
Tauri star GM Aur ({\it Schneider et al.\@}, 2003). The top left panel
shows the scattered light image and the top left a model. The lower
panel shows the SED. The dark line is a model for the stellar
photosphere emission. The white line shows a model combining the stellar
emission and the disk excess emission, with disk model parameters taken
from scattered light results. For the photometry references, see {\it
Schneider et al.\@} (2003).}
\end{figure}

Scattered light images do not generally provide information on the inner
part of the disk. In addition to IRAS data, high-quality mid-infrared spectra
from the {\it Spitzer Space Telescope} are becoming available for many
YSO disks. Scattered light images reduce many of the degrees of
freedom in modeling disk spectra and SEDs, leading directly to more
robust SED models. Future combined studies of multiwavelength disk
images, continuum spectra, and SEDs will provide critical tests for
models of disk evolution.
                   
\section{PROSPECTS FOR FUTURE ADVANCES}

\subsection{Science Goals}

The science goals for scattered-light imaging of disks around YSOs
center on gaining a better understanding of disk structure, disk
evolution, and dust evolution in the context of planet formation.

A key goal is to determine an evolutionary sequence for dust,
especially dust growth and changes in shape and composition. One
manifestation of dust evolution is to change the dust opacity law. Work
in HV Tau C and HH 30 has shown that the opacity law can be determined
from HST images in the optical and near-infrared using current modeling
codes. Extending this work is simply a matter of obtaining the relevant
observations and ``cranking the handle''. Observations of thermal
emission at longer wavelengths will help to restrict the degeneracies in
the scattered light modeling. We would expect to find disks with
different opacity laws, some with near-ISM opacity laws like HV Tau C
and others with less opacity laws like HH 30. The science will come from
comparing the opacity laws with models for dust evolution and from
correlating the opacity laws with system properties such as age,
accretion activity, disk mass, and stellar multiplicity.

The evidence for dust settling discussed in section 5.1 is as yet
tentative. To advance, we need to be able to confidently compare the
equivalent temperatures derived from the dust scale heights to the real
temperatures in the disk or, more generally, the vertical distributions
of the dust and the gas. This will require detailed calculations of the
structure of the disk that solve for the vertical structure under the
assumption of thermal equilibrium but allow for uncertainties in the
radial structure. Again, including mid-infrared and millimeter data will
be very useful.

Of course, we are ultimately interested in detecting young planets in
disks, and especially those planets that are still accreting. Direct
detection will be difficult (see the chapters by {\it Beuzit et al.\@} and
{\it Beichman et al.\@}), but we can hope to observe the gap cleared by a
planet as it grows. At the moment, these are most clearly seen in
infrared spectra, and the contribution of scattered light images is
mainly to reduce the number of degrees of freedom in the disk geometry
and inclination and thereby improve the reliability of the models for
the infrared emission.

We would like to understand the transition between optically-thick
protoplanetary disks and optically-thin debris disks in order to
determine the timescale over which planetesimals and gas -- the building
blocks of planet formation -- are present. Advances are being made in
this field with {\it Spitzer} imaging and spectroscopy. Scattered light
imaging can contribute because small amounts of dust can potentially be
detected even at radii at which infrared detections are difficult,
although the realization of this goal will require advances in
high-contrast imaging.

\subsection{Instrument Advances}

The biggest instrumental advance for the study of YSO disks will be
the advent of the Atacama Large Millimeter Array (ALMA). With spatial
resolution surpassing that of HST, very high resolution spectroscopy for
chemical and kinematic studies, and the sensitivity to study the disks
of low-mass stars as far away as Orion, ALMA will have a major impact on
the field. However, even in the era of ALMA, scattered light imaging
will still make significant contributions to disk studies. First,
scattered light traces the surface where stellar photons deliver energy to
the disk. It will still be necessary to characterize this interaction
region if the disk temperature structure, and thus its chemical nature,
is to be understood. Secondly, while millimeter wavelengths probe large
particle sizes in the dust population, scattered light imaging provides
information on the small particles. Both are needed to provide a full
picture of disk grain properties and their time evolution. The
combination of millimeter maps and scattered light images at comparable
resolutions will be a powerful synthesis for disk science efforts.

The imaging performance of large groundbased telescopes can be expected
to continue its evolution. Particularly important will be new extreme
adaptive optics instruments at the Gemini and VLT observatories.
Their improved contrast performance should enable additional detections
of YSO disks in the near- and mid-infrared. The maturation and
increasing application of differential imaging polarimetry should also
yield exciting results. Several concepts for extremely large (D $\sim$
30 m) telescopes are now being studied. When realized (maybe not until
after {\it Protostars and Planets VI}), these facilities will provide a
threefold advance in spatial resolution. Higher resolution images will
improve our knowledge of all aspects of disk structure. The inner holes
in systems such as TW Hya and perhaps GM Aur should be resolvable. An
exciting possibility is the detection of hot young planets near to and
perhaps even within YSO disks and characterization of their dynamical
interactions.

Among future space missions in design and development, two will provide
important capabilities for disk scattered light imaging. The NASA/ESA
James Webb Space Telescope will have a 6.5 m primary mirror and operate
from 1 to 28 $\mu$m. It will be a superb telescope
for imaging disks in mid-infrared scattered light and will provide
roughly 0.2\asec resolution. In the near-infrared, it will offer almost
three times better resolution than HST, but is unlikely to provide
improved contrast. From the point of view of high contrast imaging,
the Terrestrial Planet Finder Coronagraph mission would be extremely
exciting, as it would be able to detect scattered light from disks as
tenuous as our own solar system's zodiacal light. However, as this
misssion may be more than a decade away, several groups are proposing
smaller coronagraphic space telescopes that might be realized sooner.

While waiting for these future developments, the expanded application of
existing scattered light imaging capabilities (AO and mid-IR imaging
from large ground-based telescopes and high-contrast imaging with HST)
should continue unabated.

\subsection{Modeling Advances}

The new observations described in this review and the observation
advances outlines above suggest a wealth of detailed data on
circumstellar disks will become available. What will this data demand
from the codes and models?

Future codes will need to produce high-resolution images and integrated
spectra at wavelengths stretching from the optical to the millimeter.
They should be able to model three-dimensional distributions of sources
and opacity and should incorporate accurate dust scattering phase
functions, polarization, and the effects of aligned grains. Evidence for
dust growth and sedimentation requires that codes no longer restrict
themselves to homogenoeus dust properties, but must be able to treat
multiple dust species with different spatial distributions. They will
probably solve for radiation transfer using Monte Carlo techniques in
optically thin regions and the diffusion approximation in optically
thick regions. Computers are expected to become increasingly parallel in
the future. Many codes can already run in parallel, but those that
cannot will need to be modified to do so. In this context, Monte Carlo
algorithms have the advantage over classical algorithms as they often
have natural parallelism. Many radiation transfer codes now have the
capability to determine the density structure from given disk physics or
incorporate density structures from dynamical simulations. These will be
increasingly useful in providing detailed tests of disk structure. On
the other hand, parameterized density models will continue to
be a useful tool for mitigating our incomplete knowledge of disk
physics.

The combination of future data and future codes will allow us to study
dust properties, dust settling, disk structure, disk-planet
interactions, accretion, and disk evolution.
In the longer term, advances in techniques for numerical simulations
coupled with increases in computing power and parallel processing make
accurate radiation hydrodynamic simulations of disk formation and
evolution a distant but realistic goal.

Perhaps the most important near-term work is to apply scattered light
modeling techniques across already extant disk image datasets.
Observers have provided a significant number of new, high-quality
images, but the corresponding modeling efforts have not kept pace.

%
%
%
%
%

\acknowledgements We thank an anonymous referee for comments that helped
to improve this chapter. This work was partially supported by the Centro
de Radioastronom{\'\i}a y Astrof{\'\i}sica of the Universidad Nacional
Aut\'onoma de M\'exico, HST GO program 9424 funding to the Jet
Propulsion Laboratory of the California Insitute of Technology, and the
Programme National de Physique Stellaire (PNPS) of CNRS/INSU, France.

\end{document}